\documentclass[10pt]{iopart}
\usepackage{iopams}
\usepackage{amssymb,amsfonts,mathrsfs,setstack}
\usepackage[dvips]{graphicx}
\begin{document}

\newcommand{\B}{\underline{B}}
\newcommand{\N}{\underline{N}}
\newcommand{\Bb}{\bar{B}}
\newcommand{\W}[1]{\mathcal{W}(#1)}
\newcommand{\Hilb}{\mathscr{H}}
\newcommand{\I}{\mathbb{I}}
\newcommand{\n}{\hat{n}}
\newcommand{\ntot}{\n_{\small{\mathrm{tot}}}}
\newcommand{\un}[1]{\mathfrak{u}(#1)}
\newcommand{\su}[1]{\mathfrak{su}(#1)}
\newcommand{\ab}[1]{a^{\phantom{\dag}}_{#1}}
\newcommand{\ad}[1]{a^{\dag}_{#1}}
\newcommand{\alg}[1]{\mathcal{#1}}
\newcommand{\bb}[1]{b^{\phantom{\dag}}_{#1}}
\newcommand{\bd}[1]{b^{\dag}_{#1}}
\newcommand{\LOP}[1]{\underline{#1}}
\newcommand{\LOPF}[1]{\bar{#1}}
\newcommand{\indotta}{\widehat{\mathrm{JS}}}

\title[An Algebraic Approach to Linear-Optical Schemes for Deterministic QC]{An Algebraic Approach to Linear-Optical Schemes for Deterministic Quantum Computing}
\date{\today}

\author{Paolo Aniello\dag\ddag\ and Ruben Coen Cagli\dag\ddag
\footnote[3]{coen@na.infn.it}
}

\address{\dag\ Dipartimento di Scienze Fisiche,\\
		Universit\`a degli Studi di Napoli 'Federico II',\\
             	Complesso Universitario di M. S. Angelo, Napoli, Italy\\}

\address{\ddag\ Istituto Nazionale di Fisica Nucleare,\\
		Sezione di Napoli}

\begin{abstract}
Linear-Optical Passive (LOP) devices and photon counters are sufficient to
implement universal quantum computation with single photons, and
particular schemes have already been proposed.
In this paper we discuss the link between the algebraic
structure of LOP transformations and quantum computing.
We first show how to decompose the Fock space of $N$ optical modes in
finite-dimensional subspaces that are suitable for encoding strings of
qubits and invariant under LOP transformations (these subspaces
are related to the
spaces of irreducible unitary representations of U($N$)). Next we
show how
to design in algorithmic fashion
 LOP circuits which implement any quantum circuit deterministically. We
also present some simple examples, such as the circuits implementing a
CNOT gate and a Bell-State Generator/Analyzer.
\end{abstract}






\section{\label{sec1}
         Introduction}
Since the early times of quantum computing (QC) and quantum information
 processing (QIP), several proposals of implementation schemes
have come from the field of quantum optics ~\cite{Nielsen}.

Quite recently, see ~\cite{klm1}, it has been shown that, at least in principle,
\emph{scalable-nondeterministic} quantum computation can be achieved by
linear-optical passive (LOP) devices, exploiting also the nonlinearity
introduced by conditional measurements ~\cite{scheel,lapaire,clausen}.
This has lead to several schemes ~\cite{pjf1,ral1,ral2,ral3,knill,ral4,giorgi} and experimental
demonstrations ~\cite{pjf2,san1,obra,pjf3,gas,zhao,san2} of quantum gates
and circuits, which exploit different ways to encode a \emph{single qubit} by a \emph{single photon}:
the qubit basis states can be encoded by the vacuum
and the one-photon Fock states $\{ |0\rangle,|1\rangle\}$ of a given mode of the
quantized e.m. field  ~\cite{milburn}, or by two orthogonal polarization states of a photon
$\{ |H\rangle,|V\rangle\}$, or by the one-photon Fock states of a two-mode
optical system $\{ |01\rangle,|10\rangle\}$.

On the other hand, the possibility of \emph{deterministic-nonscalable}
linear-optical quantum computing has been also pointed out, see ~\cite{cerf,kurt}, and
fundamental gates have been experimentally tested ~\cite{fiore1,fiore2}.
These works rely on \emph{single-photon multi-qubit} (SPMQ)
encoding schemes, namely schemes that allow to encode, say, $k$ qubits
by a single photon, by introducing $2^{k}-1$ vacuum optical modes.

The aim of the present paper is twofold: first, in \S \ref{sec2}, we show that the SPMQ
encoding stems in a natural way from simple general features of the fundamental
algebraic objects associated with the description of the quantized e.m. field and of LOP devices;
then, in \S \ref{sec3}, we make use of this result to introduce a simple algorithmic procedure which,
for any given quantum computation, allows
		to design the LOP circuit that implements it deterministically. We also briefly discuss the issue of
scalability. Eventually, in \S \ref{sec4}, we
present some basic examples of deterministic LOP circuits.

\section{\label{sec2}
         Algebraic tools}
In this section we introduce the  algebraic formalism necessary to describe
a quantum system of $N$ optical modes, and the LOP devices acting on it;
our aim is to highlight the algebraic structure
underlying the implementation of QIP and
QC by linear optics.

	\subsection{\label{sec2.1}
	            Heisenberg-Weyl algebra and Fock space}

	We start by briefly recalling the algebraic description of the
	physical system of a single quantum optical mode, i.e. a vibrational
mode of fixed frequency
	of the quantized e.m. field, that is formally equivalent to a quantum
harmonic oscillator.

The fundamental object of this description is the \emph{Heisenberg-Weyl} algebra $\W{1}$, namely the complex Lie
algebra with generators $\{ \mathtt{a},\mathtt{a}^{\star},\mathtt{e}\}$ satisfying the relations:
 \begin{eqnarray}\label{commuta1}
	\lbrack \mathtt{a}, \mathtt{a}^{\star}\rbrack & = &\mathtt{e}  \quad ,{} \nonumber\\
	  \lbrack \mathtt{a}, \mathtt{e}\rbrack & = & 0 \,  =  \lbrack \mathtt{a}^{\star}, \mathtt{e}\rbrack \quad , {}
 \end{eqnarray}
	where $\lbrack \cdot , \cdot \rbrack $ denotes the Lie bracket.
Notice that $\W{1}$ is a $\star$-algebra, since it is endowed with the involution $\star$, namely the antilinear
map determined by:
\begin{equation}
\mathtt{a}\mapsto\mathtt{a}^{\star}\quad ,\quad \mathtt{a}^{\star}\mapsto(\mathtt{a}^{\star})^{\star}=\mathtt{a} \quad , \quad
 \mathtt{e}\mapsto\mathtt{e}^{\star}=\mathtt{e} \quad .
\end{equation}
The algebra $\W{1}$ admits a remarkable realization --- realization which will be still denoted
by $\W{1}$ in the following --- as an algebra of
operators in an infinite-dimensional Hilbert
 space $\Hilb_{F}$ (with the Lie bracket realized by
the commutator):
\begin{equation}\label{realization}
\mathtt{a}\mapsto a\quad ,\quad \mathtt{a}^{\star} \mapsto a^{\dag}
\quad , \quad\mathtt{e}\mapsto \I \quad ,
\end{equation}
where $a^{\dag}$ is the Hilbert space adjoint of $a$ and  $\I$ is the
identity operator. In fact, given an orthonormal basis
$\{|n\rangle\}_{n=0}^{\infty}\subset\Hilb_{F}$, one can define the
\emph{annihilation operator} $a$ by
\begin{equation}\label{annihilation}
a|n\rangle = \sqrt{n}|n-1\rangle\quad \mathrm{for}
	 \quad n\geq 1\quad , \quad a\, |0\rangle = 0 \, .
\end{equation}
It follows that the \emph{creation operator} $a^{\dag}$ satisfies:
\begin{equation}
a^{\dag}|n\rangle = \sqrt{n+1}|n+1\rangle \quad n=0,1,\ldots\, ;
\end{equation}
moreover, one can define the \emph{number operator} $\hat{n}\doteq a^{\dag}a$,
\begin{equation}
\hat{n}|n\rangle = n|n\rangle, \quad n=0,1,\ldots \, ,
\end{equation}
which is a positive self-adjoint operator.\\
Then, one can easily verify that
\begin{equation}\label{commutation1}
\lbrack a , a^{\dag} \rbrack = \I \quad ,
\end{equation}
 hence, $\{a, a^{\dag},\I\}$ generate indeed a realization
of the Heisenberg-Weyl algebra. The infinite-dimensional Hilbert space $\Hilb_{F}$
--- endowed with the orthonormal basis $\{|n\rangle\}_{n=0}^{\infty}$ and
the associated operator algebra generated by $\{a, a^{\dag},\I\}$  ---
is called \emph{one-mode} (bosonic) \emph{Fock space}.

The operator realization of $\W{1}$ can also be introduced in a more
abstract way. In fact, let $a$ be a linear operator in a (complex separable) Hilbert space
 $\Hilb_{F}$ satisfying
relation (\ref{commutation1}). It is then possible to prove that, if,
 in addition, $\{a, a^{\dag}\}$  is an irreducible set of operators
(i.e. any non-trivial linear span which is invariant under the action of $a$
and $a^{\dag}$ must be dense in $\Hilb_{F}$) and a technical condition
concerning the operator $\hat{n}\doteq a^{\dag}a$ is verified \cite{dixmier},
the Hilbert space $\Hilb_{F}$ must be infinite-dimensional and the operators
$a,a^{\dag}$ are unitarily equivalent to the standard annihilation and creation
operators of the harmonic oscillator in $L^{2}(\mathbb{R})$; moreover,
the unitary operator that generates this unitary equivalence is uniquely
defined up to an arbitrary  phase  factor. This is one of the formulations
of the Stone-von Neumann theorem on  canonical commutation relations
\cite{stone} \cite{von neumann}. As a consequence, there is an
orthonormal basis $\{|n\rangle\}_{n=0}^{\infty}$ (defined uniquely up
to an overall phase factor) in $\Hilb_{F}$ such that (\ref{annihilation}) is satisfied;
hence, one recovers the previous definition of the operator realization of
$\W{1}$.
\\
The operators $a,a^{\dag}$ defined in such a abstract way play respectively
the role of annihilation and creation operators of the quantized e.m. field,
and $\hat{n}$ is then the photon-number operator.
\\
Finally, we notice that $\Hilb_{F}$ can be decomposed
in a natural way as the direct
sum of subspaces characterized by a given number of photons:
		\begin{equation}\label{directsum}
		 \Hilb_{F} = \bigoplus_{n=0}^{\infty}\Hilb_{n} \quad
		,\quad\Hilb_{n}
		 =\mathrm{span}\{ |n\rangle \} = \{ \alpha|n\rangle\, :\, \alpha\in\mathbb{C} \} \quad.
        \end{equation}

This formalism can be extended to the general case of $N$ optical modes.
The fundamental object becomes
the algebra $\W{N}$, which is realized as the subspace of the linear space $\W{1}^{\otimes N}$
generated by the basis elements $\{\ab{i},\ad{i},\I\}_{i=1}^{N}$, where:
\begin{equation}
\ab{i}=\I\otimes\cdots \otimes\I\otimes\overbrace{a}^{i}\otimes\I\otimes \cdots \otimes\I \quad ,
\end{equation}
satisfying the canonical commutation  relations:
\begin{equation}\label{commutaN}
	\lbrack \ab{i}, \ad{j}\rbrack =  \delta_{ij}\I \quad ,\qquad i,j=1,2,\ldots N \quad .
\end{equation}
The Hilbert space of this
 realization is the $N$\emph{-mode} (bosonic) \emph{Fock
space} $\Hilb^{(N)}_{F}\doteq\Hilb_{F}^{\otimes N}$
endowed with the orthonormal basis $\{|n_{1}\ldots n_{N}\rangle\, : \, n_{1},\ldots ,n_{N}=0,1,\ldots\}$, with:
\begin{equation}\label{stati}
	|n_{1},\ldots n_{N}\rangle = \left(
	\prod_{i=1}^{N}\frac{(\ad{i})^{n_{i}}}{\sqrt{n_{i}!}}\right)
	|\mathbf{0}\rangle \quad .
\end{equation}
Notice in passing that, if $N\geq 2$, the set of operators
$\{\ab{i},\ad{i},\I\}$, for any $i\in\{1,\ldots N\}$, generates now
a \emph{reducible} realization of $\W{1} $.
\\
Notice also that, on the other hand, the operators $\{\ab{i},\ad{i}\}_{i=1}^{N}$
indeed form an irreducible set, and, according to the Stone-von Neumann theorem,
any other irreducible set of operators $\{\bb{i},\bd{i}\}_{i=1}^{N}$
satisfying the canonical commutation relations (\ref{commutaN}) must be related
to the set $\{\ab{i},\ad{i}\}_{i=1}^{N}$ by a unitary equivalence; precisely,
there is a unitary operator $U$ in $\Hilb_{F}^{(N)}$,
\emph{uniquely defined up to an arbitrary phase factor}, such that:
\begin{equation}
\ab{i}=U\,\bb{i}\,U^{\dag}\quad ,\quad \ad{i}=U\,\bd{i}\,U^{\dag}\, .
\end{equation}
$\quad$\\
The Hilbert space  $\Hilb^{(N)}_{F}$ can be decomposed as:
\begin{equation}\label{directsumN}
		 \Hilb_{F}^{(N)} = \bigoplus_{n=0}^{\infty}
		\Hilb_{n}^{(N)} \quad ,
\end{equation}
where $\Hilb_{n}^{(N)}$ is the subspace characterized by a given number $n$ of photons,
i.e.
\begin{equation}
		\Hilb_{n}^{(N)} = \mathrm{span}\{|n_{1}\ldots n_{N}\rangle\, : \, n_{1}+\cdots +n_{N}=n \} \quad .
\end{equation}
These subspaces are the eigenspaces of the 'total number of photons operator', i.e. of the
positive self-adjoint operator
\begin{equation}
\n = \sum_{i=1}^{N}\ad{i}\ab{i} \quad .
\end{equation}
Decomposition (\ref{directsumN}) plays a central role in understanding linear-optical quantum computing,
and we will characterize it in the following by means of the bosonic realization of
the Lie algebra  $\un{N}$ of U($N$)
that is obtained via the Jordan-Schwinger map.

	\subsection{\label{sec2.2}
	            The Jordan-Schwinger map}
	   In QIP and QC one always deals with finite-dimensional Hilbert spaces;
	indeed, information is represented by
	\emph{words} over some finite \emph{alphabet}, namely by finite strings of symbols,
	and in the quantum domain these symbols
	are realized by \emph{qubits} or, more in general, by \emph{qudits}.
	From a mathematical point of view, a
	\emph{logical} qudit is a vector in a $d$-dimensional
	abstract Hilbert space, and strings of symbols are obtained through the tensor product structure.

	On the other hand, when dealing with quantum optical systems one
	works with the \emph{infinite-dimensional} Fock
	space. Nevertheless, we will show that by means of the
	Jordan-Schwinger (J-S)
	map, one can single out in a natural way suitable \emph{finite-dimensional}
	subspaces that allow
	to encode qudits and to represent the appropriate class of
	transformations that allow to 'move within' these
	subspaces, namely to represent the action of quantum logic gates.

	The general formulation
	of the J-S map ~\cite{jordan, schw, marmo, wun} gives a simple procedure
	allowing to
	obtain the so called \emph{bosonic realization} of a Lie
	algebra.
	Consider  an operator realization of the
	$\mathfrak{gl}(N)$
	algebra, with basis elements given by $\{ d_{ij}\doteq\ad{i}\ab{j} \, :\, i,j=1,\ldots N\}$:
	\begin{equation}\label{dij}
	\lbrack d_{ij} , d_{mn} \rbrack = \delta_{jm}d_{in}-\delta_{ni}d_{mj}
	\quad .
	\end{equation}
	Now, consider a $M$-dimensional matrix Lie algebra $\alg{A}$,
	and a basis $\{A_{k}\}_{k=1}^{M}$ of $\alg{A}$ of, say, $N\times N$  matrices
	($M\leq N^{2}$) \footnote[3]{Recall that, by Ado's theorem,
		any finite-dimensional complex Lie algebra is isomorphic to a subalgebra of
		$\mathfrak{gl}(N)\equiv\mathfrak{gl}(N,\mathbb{C})$ for some $N$.}.
	Due to the properties of the operator realization of $\W{N}$,
	one can define the linear operators
	\begin{equation}\label{generatori}
     Q_{k} = \sum_{ij}(A_{k})_{ij}\, d_{ij}= \sum_{ij}\ad{i}(A_{k})_{ij}\ab{j}
	\quad ,\, k=1,\ldots M\, ,
    \end{equation}
	associated with the natural action of the matrices
	$\{A_{k}\}_{k=1}^{M}$ on the linear span of $\ab{1},\ldots ,\ab{N}$.
	The linear operators $\{Q_{k}\}_{k=1}^{M}$
	provide a basis of the $N$-dimensional
	'bosonic realization' $\alg{Q}$ of $\alg{A}$ since,
	as the reader may verify using relations~{(\ref{dij}),} the
	operators $\{ Q_{k}\}_{k=1}^{M} $ preserve the commutation rules
	of the basis matrices $\{ A_{k}\}_{k=1}^{M} $:
	\begin{equation}\label{commutation}
     \lbrack Q_{k},Q_{l}\rbrack =
	  \sum_{ij}\left( \lbrack A_{k},A_{l}\rbrack\right)^{\phantom{\dag}}_{ij}\, d_{ij}
	\quad , \quad k,l=1,\ldots ,M \, .
    \end{equation}
	The one-to-one correspondence $A_{k}\mapsto Q_{k}\, ,\, k=1,\ldots ,M$ ---
	\emph{extended by linearity} --- is the J-S map:
	$\mathrm{JS}\, :\, \alg{A}\rightarrow \alg{Q}$.

	As a first example, let us construct the bosonic realization of the Lie algebra of the
	group U(2). To this aim, we have to
	consider the composite system of two optical
	modes. Its Fock space $\Hilb_{F}^{(2)}$ can be decomposed as in (\ref{directsumN}), with:
	       \begin{equation}\label{subspace}
	        \Hilb_{n}^{(2)} =\mathrm{span}\{ |n,0\rangle , |n-1,1\rangle , \ldots ,|0,n\rangle\} \quad.
	       \end{equation}
	$\quad$\\
	Making use of the J-S map, it can be shown that these are the
	spaces of the irreducible unitary representations of the
	group U(2) (or, also, of the group SU(2)).
	\\
	Indeed, consider $\alg{A}=\un{2}$ \footnote[3]{We follow the physicists' convention
			for the Lie algebra $\un{N}$;
			as a consequence an imaginary unit will appear
			in the argument of the exponential map
			from the algebra onto the group U($N$).} and its
	matrix realization generated by the $2\times 2$ identity matrix Id and
	by the Pauli matrices
	$\sigma_{1},\sigma_{2},\sigma_{3}$;  the generators of
	the bosonic
	realization of $\un{2}$ are then obtained applying formula (\ref{generatori}):
	       \begin{eqnarray}\label{J}
		J_{1}\doteq \mathrm{JS}\Big(\frac{1}{2}\sigma_{1}\Big) & = &  \frac{1}{2}\left( \begin{array}{cc}
		                                  \ad{1} & \ad{2} \end{array}\right)
		                           \left( \begin{array}{cc}
		                                 0 & 1\\ 1 & 0 \end{array}\right)
		                             \left( \begin{array}{c}
		                                   \ab{1} \\ \ab{2} \end{array}\right)  \nonumber \\
		    & = & \frac{1}{2}(\ab{1}\ad{2}+\ad{1}\ab{2}) \quad ,\nonumber \\
		J_{2}\doteq \mathrm{JS}\Big(\frac{1}{2}\sigma_{2}\Big)  & = &  \frac{1}{2}\left( \begin{array}{cc}
		                                   \ad{1} & \ad{2} \end{array}\right)
		                            \left( \begin{array}{cc}
		                                  0 & i\\ -i & 0 \end{array}\right)
		                             \left( \begin{array}{c}
		                                   \ab{1} \\ \ab{2} \end{array}\right)  \nonumber \\
		    & = & \frac{i}{2}(\ad{1}\ab{2}-\ab{1}\ad{2})\quad ,\nonumber \\
		J_{3}\doteq \mathrm{JS}\Big(\frac{1}{2}\sigma_{3}\Big) & = & \frac{1}{2} \left( \begin{array}{cc}
	                                        \ad{1} & \ad{2} \end{array}\right)
	                                 \left( \begin{array}{cc}
	                                       1 & 0\\ 0 & -1 \end{array}\right)
	                                    \left( \begin{array}{c}
		                                  \ab{1} \\ \ab{2} \end{array}\right) \nonumber \\
		    & = &\frac{1}{2}(\ad{1} \ab{1}-\ad{2} \ab{2}) \quad ,\nonumber  \\
		\n \doteq \mathrm{JS}(\mathrm{Id}) & = & \left( \begin{array}{cc}
	                                        \ad{1} & \ad{2} \end{array}\right)
	                                 \left( \begin{array}{cc}
	                                       1 & 0\\ 0 & 1 \end{array}\right)
	                                    \left( \begin{array}{c}
		                                  \ab{1} \\ \ab{2} \end{array}\right) \nonumber \\
		    & = & (\ad{1} \ab{1}+\ad{2} \ab{2}) \quad ,
		   \end{eqnarray}
	where one can verify that the standard 'angular momentum commutation rules' are satisfied:
		   \begin{equation}
		    \lbrack J_{i},J_{j} \rbrack = i\epsilon_{ijk} J_{k} \quad,
		   \end{equation}
	with $\epsilon_{ijk}$ denoting the Levi Civita tensor, and $\lbrack J_{k},\n\rbrack=0$.
	The action of the operator $J_{3}$ and of the Casimir operator
	\begin{equation} \mathbf{J}^{2}={J}_{1}^{2}+{J}_{2}^{2}+{J}_{3}^{2}=
	\frac{\n}{2}\left( \frac{\n}{2}+1\right)
	\end{equation}
	on the Fock states, i.e.:
		   \begin{eqnarray}
		    \mathbf{J}^{2}|n_{1}n_{2}\rangle & = &
		    \frac{n_{1}+n_{2}}{2}\left( \frac{n_{1}+n_{2}}{2}+1\right) |n_{1}n_{2}\rangle  \quad ,\\
		    J_{3}|n_{1}n_{2}\rangle & = & \frac{n_{1}-n_{2}}{2}|n_{1}n_{2}\rangle \quad,
		   \end{eqnarray}
	gives rise to a relabelling of these states as
	eigenstates of the abstract angular momentum:
		   \begin{equation}\label{relabel}
		    |n_{1}n_{2}\rangle = |j,l\rangle
		   \end{equation}
	where
	\begin{equation}
	j=\frac{n_{1}+n_{2}}{2}\quad \mathrm{and}\quad
	l=\frac{n_{1}-n_{2}}{2} \quad ,
	\end{equation}
	so that
	the subspaces $\{\Hilb_{n}^{(2)}\}_{n=0}^{\infty}$ in (\ref{subspace}) are the
	spaces of the ($2j+1$)-dimensional unitary
	irreducible representation of the U(2) group:
		   \begin{equation}
		   \label{subspace2}
		    \Hilb_{n}^{(2)} =\mathrm{span}\{ |j,j\rangle ,
            |j,j-1\rangle , \ldots |j,-j\rangle\}\, , \, j=\frac{1}{2}n \, ,
		   \end{equation}
	with the index $j$ identifying these subspaces, and the index $l=-j,\ldots j$
		  labelling the standard basis vectors in each subspace.

	The generalization to the multimode case,
	with $N\geq 3$, can be outlined as follows.
	By the J-S map, the $\un{N}$ generators can be realized as linear superpositions
	of the operators $\{\ad{i}\ab{j}\}_{i,j=1}^{N}\subset\W{1}^{\otimes N}$.
	This makes it clear that the subspaces $\Hilb_{n}^{(N)}$ of the $N$-mode
	Fock space $\Hilb_{F}^{(N)}$ are invariant subspaces for the bosonic realization
	of the $\un{N}$ generators: indeed, the action of the operators $\{\ad{i}\ab{j}\}_{i,j=1}^{N}$
	\emph{preserves the total number of photons}. It follows that the $n$-photon space
	$\Hilb_{n}^{(N)}$ can be decomposed as an orthogonal sum of spaces of irreducible
	unitary representations of U($N$). Actually, using the formalism of the 'highest weights'
	(see, for instance, ~\cite{simon}), one can prove that, as in the special case of U(2), the $n$-photon space
	$\Hilb_{n}^{(N)}$ is the space of an irreducible unitary representation of U($N$) (hence,
	the mentioned orthogonal sum contains only one term).
	One can show, moreover, that, for $N\geq 3$, not all the irreducible
	representations of U($N$) can be realized in such a way; for instance, for $N=3$,
	only the representations of dimension $\frac{1}{2}(n+1)(n+2)$, $n=0,1,\ldots$, can be
	realized, while it is well known that the dimension of the irreducible unitary representations
	of U(3) is given by the general formula:
	\begin{equation}
	d=\frac{1}{2}(n_{1}+1)(n_{2}+1)(n_{1}+n_{2}+2)\quad, \quad n_{1},n_{2}=0,1,\ldots\quad .
	\end{equation}
	However, in what follows we will essentially deal
	with the \emph{definitory} (or \emph{fundamental}) representation of U($N$),
	whose Hilbert space is the single-photon $N$-mode
	space $\Hilb_{1}^{(N)}$.

	The characterization of $\Hilb_{1}^{(N)}$ that we have given fits with the abstract definition
	of the Hilbert space of a qu$N$it: a $N$-dimensional Hilbert space endowed with the fundamental
	representation of U($N$) acting on it.	In the following, we will be
	specifically interested in the values of $N$ given by
	$N=2^{k}\, , \, k=1,2,\ldots\,$. For these values of $N$, the following Hilbert space
	isomorphisms hold:
	\begin{equation}\label{kQubitSpace}
	\Hilb_{1}^{(2^{k})}\cong (\mathbb{C}^{2})^{\otimes k} \cong
	\Hilb_{1}^{(2)}\otimes\cdots\otimes \Hilb_{1}^{(2)} \quad ,
	\end{equation}
	with $(\mathbb{C}^{2})^{\otimes k}$ regarded as an abstract qu$(N=2^{k})$it, or, equivalently,
	as a $k$-qubit. However we stress the following points:
	\begin{itemize}
		\item the Hilbert spaces respectively on the l.h.s. and on the r.h.s.
		of (\ref{kQubitSpace}) --- though mathematically isomorphic ---
		have, for $k\geq 2$, a different physical meaning, since the former is a
		single-photon space while the latter is a $k$-photon space;
		\item this physical content has its mathematical counterpart in the fact that,
		for $k\geq 2$, the Fock space $\Hilb_{F}^{(2^{k})}$ is endowed with an irreducible operator
		realization of the algebra $\W{2^{k}}$, while the space $(\Hilb_{F}^{(2)})^{\otimes k}$
		is endowed just with $k$ reducible operator realizations of $\W{2}$;
		\item accordingly, using the J-S map, one can endow $\Hilb_{1}^{(2^{k})}$
		with the fundamental representation of U(2$^{k}$), while, by the same procedure,
		only the fundamental representation of:
		\begin{equation}
		\mathrm{U}(2)\otimes\cdots\otimes \mathrm{U}(2) \subsetneq \mathrm{U}(2^{k})\quad (k\geq 2)
		\end{equation}
		can be obtained (namely, $(\Hilb_{1}^{(2)})^{\otimes k}$ is the space of $k$ qubits
		which 'do not interact').
	\end{itemize}
	The previous observations are the basis of the SPMQ encoding and of the use of LOP
	transformation for the implementation of logic gates.

	\subsection{\label{sec2.3}
	            LOP transformations}
	We will now move to the description of the class
	of optical transformations
	which enable to implement the \emph{elaboration}
	of quantum information, namely
	logic gates on qu$N$its.

\begin{figure}[!h]
	   \begin{center}
	    \includegraphics[width=0.15\textwidth]{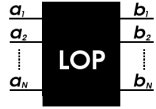}
	    \caption{\label{fig1} Generic LOP multiport.
	    Horizontal lines represent optical modes entering
	    and leaving the device, from left to right. }
	   \end{center}
	  \end{figure}

    The linear-optical \emph{passive} (LOP)
    transformations are defined as the class of linear
    transformations that act on the system of $N$ optical modes
    --- i.e. of linear transformations in span$\{\ab{1}\ldots\ab{N}\}$ ---
    leaving unchanged the total number of photons in the process; a generic LOP device
    is usually depicted as $2N$-port,
    namely a black box with $N$ inputs and $N$ outputs (Fig.~\ref{fig1}), respectively corresponding
    to the field operators $\{\ab{i}\}_{i=1}^{N}$ and $\{\bb{i}\}_{i=1}^{N}$.

	  The property of photon-number conservation is expressed by the condition:
	\begin{equation}\label{LOP}
          \sum_{i}\ad{i}\ab{i} = \sum_{i}\bd{i}\bb{i} \quad .
         \end{equation}
	A simple calculation shows that this condition is sufficient to guarantee that the canonical
	commutation relations (\ref{commutaN}) still hold for the operators $\{\bb{i},\bd{i},\I\}_{i=1}^{N}$,
	which thus form another basis for the realization of $\W{N}$, and can indeed be interpreted as the
	field operators of the output modes.\\
	In fact, denoting by $\LOP{U}$ the matrix representing the LOP transformation:
	\begin{equation}\label{matriceTrasf}
          \bb{i} = \sum_{j}\LOP{U}_{ij}\ab{j}\quad ,
        \end{equation}
	from condition (\ref{LOP}) one can easily prove that
	\begin{eqnarray}
          \LOP{U}\,\,\LOP{U}^{\dag} = \LOP{U}^{\dag}\,\LOP{U} = \mathrm{Id}\quad ,
         \end{eqnarray}
	where Id is the identity matrix. Hence, with any LOP transformation it can be naturally
	associated a \emph{unitary matrix} $\LOP{U}$; conversely, any unitary matrix defines
	a LOP transformation. Thus, there is a one-to-one correspondence between LOP $2N$-ports and the elements
	of the group U($N$).

	On the other hand, formula (\ref{matriceTrasf}) implies that 
	any invariant linear span for the operators $\{\bb{i},\bd{i}\}_{i=1}^{N}$
	must be invariant for the operators $\{\ab{i},\ad{i}\}_{i=1}^{N}$ as well,
	and this in turn means that the output field operators
	$\{\bb{i},\bd{i}\}_{i=1}^{N}$ form another irreducible set.
	Thus, as we have already pointed out in \S \ref{sec2.1},
	by the Stone-von Neumann theorem they must be
	unitarily equivalent to the operators $\{\ab{i},\ad{i}\}_{i=1}^{N}$, i.e.
	\begin{equation}
          \bb{i} = U^{\dag}\ab{i}U \quad , \quad i=1,\ldots, N\quad ,
         \end{equation}
	where $U$ is a unitary operator in $\Hilb_{F}^{(N)}$ uniquely defined up to an arbitrary
	phase factor.

	Observe that, as a consequence of the photon-number conservation, the
	subspaces $\Hilb_{n}^{(N)}\, ,\, n=0,1,\ldots$ of $\Hilb_{F}^{(N)}$
	are invariant subspaces for the operator $U$. In fact,
	using the definition of $\n$ and condition  (\ref{LOP}),
	one can easily check that the unitary operator  $U$ commutes with $\n$:
	\begin{eqnarray}
	U\, \n 			 & = & U\,\n \,U^{\dag}\,U \nonumber \\
				 & = & \left( \sum_{i}U\,\bd{i}\bb{i}\,U^{\dag} \right) U\nonumber \\
				 & = & \sum_{i}\ad{i}\ab{i} \,U \nonumber \\
				 & = & \n \, U \quad .
	\end{eqnarray}
	Then, since $\Hilb_{n}^{(N)}\, ,\, n=0,1,\ldots$, is an eigenspace of $\n$,
	it must be an invariant subspace for $U$.\\
	In particular, as $\Hilb_{0}^{(N)}=$ span$\{|\mathbf{0}\rangle\}$, we have:
	\begin{equation}\label{fasediU}
	U|\mathbf{0}\rangle = e^{i\phi}|\mathbf{0}\rangle \quad ,
	\end{equation}
	for some $\phi\in\mathbb{R}$. We will now show that one can give an explicit form
	of the operator $U$ in such a way that $e^{i\phi}=1$. To this aim, consider the following recipe:
	\begin{itemize}
	\item write the matrix $\LOP{U}$ (associated with any LOP $2N$-port) as the exponential of a matrix
	in the Lie algebra $\un{N}$:
		\begin{equation}
		\LOP{U} = e^{i\LOP{J}} \quad , \quad \LOP{J}\in\un{N}\quad ;
		\end{equation}
	\item next, via the J-S map, one can obtain a self-adjoint operator $J$:
		\begin{equation}
		J = \mathrm{JS}(\LOP{J}) = \sum_{ij}\LOP{J}_{ij}\ad{i}\ab{j} \quad ;
		\end{equation}
	\item eventually one can define a unitary operator
		\begin{equation}\label{operatoreUnitario}
		U \doteq \mathrm{exp}(i\mathrm{JS}(\LOP{J})) =
		\mathrm{exp}\big(i\sum_{ij}\LOP{J}_{ij}\ad{i}\ab{j}\big) \quad .
		\end{equation}
	\end{itemize}
	We now claim that
	\begin{enumerate}
	\item  the unitary operator $U$ verifies eq. (\ref{matriceTrasf});
	\item  the definition of $U$ does not depend on the choice of a particular
		 element $\LOP{J}$ of the algebra $\un{N}$ such that exp$(i\LOP{J})=\LOP{U}$;
	\item  the matrix $\LOPF{U}_{1}^{(N)}$ representing the operator $U$ in the one-photon subspace
		 $\Hilb_{1}^{(N)}$ of $\Hilb_{F}^{(N)}$ is $\LOP{U}$, precisely:
		 \begin{equation}\label{coincidenzaMatrici}
		 \langle 0, \ldots , 0,\overbrace{1}^{i},0,\ldots ,0|\,U\, |
		 0, \ldots , 0,\overbrace{1}^{j},0,\ldots ,0\rangle
		 = \LOP{U}_{ij} \quad .
		 \end{equation}
	\end{enumerate}
	Indeed, let $\LOP{J}$ be a matrix in $\un{N}$ and let us define the unitary operator
	$U$ by formula (\ref{operatoreUnitario}). Then, using the well known relation
	\begin{equation}
	e^{\hat{A}}\,\hat{B}\,e^{-\hat{A}} = \mathrm{exp}(\mathrm{ad}_{\hat{A}})\hat{B}
	\end{equation}
	that holds for generic linear operators $\hat{A},\hat{B}$
	(with ad$_{\hat{A}}\hat{B}\doteq\lbrack \hat{A},\hat{B}\rbrack$), and applying the canonical
	commutation relations, one easily proves that
	\begin{equation}
	U^{\dag}\ab{k}U = \sum_{l}(e^{i\LOP{J}})_{kl}\ab{l} \quad , \quad k = 1,\ldots , N \, ;
	\end{equation}
	hence, for any $\LOP{J}$ satisfying $e^{i\LOP{J}}=\LOP{U}$, the operator $U$ verifies
	eq. (\ref{matriceTrasf}). \\
	Next, we prove that the association $\LOP{U}\mapsto{U}$ defined by formula
	(\ref{operatoreUnitario}) does not depend on the choice of the matrix $\LOP{J}$
	such that $e^{i\LOP{J}}=\LOP{U}$. To this aim, observe that --- according to the
	Stone-von Neumann theorem --- the operator $U$, which verifies eq. (\ref{matriceTrasf}),
	is uniquely identified by the phase factor $e^{i\phi}$ appearing in eq. (\ref{fasediU}).
	Now, one can immediately check that, for any matrix $\LOP{J}\in\un{N}$, we have:
	\begin{equation}
	\mathrm{exp}\Big( i\sum_{k,l}\LOP{J}_{kl}\ad{k}\ab{l}\Big)|\mathbf{0}\rangle
	= |\mathbf{0}\rangle \quad ;
	\end{equation}
	hence, $e^{i\phi}=1$ for $U$ defined by formula (\ref{operatoreUnitario})
	independently on the choice of $\LOP{J}$ such that $e^{i\LOP{J}}=\LOP{U}$. This
	proves that the definition of $U$ itself does not depend on a particular choice
	of such a matrix $\LOP{J}$, namely, our second claim. \\
	Our third claim can be checked by an elementary calculation, by explicitly
	evaluating the l.h.s of eq. (\ref{coincidenzaMatrici}):
	\begin{eqnarray}\label{primoBlocco}
	 \langle \mathbf{0}|\ab{i}U\ad{j}|\mathbf{0}\rangle
					& = & \langle \mathbf{0}|U(U^{\dag}\ab{i}U)\ad{j}|\mathbf{0}\rangle \nonumber \\
					& = & \langle \mathbf{0}|\bb{i}\ad{j}|\mathbf{0}\rangle \nonumber \\
					& = & \sum_{k}\LOP{U}_{ik}\langle\mathbf{0}|\ab{k}\ad{j}|\mathbf{0}\rangle \nonumber
					\\
					& = & \LOP{U}_{ij}\quad .
	\end{eqnarray}
	(in the second line we used the fact that $U|\mathbf{0}\rangle=|\mathbf{0}\rangle \, ,\, U^{\dag}\ab{i}U=\bb{i}$).
	Summarizing, we have shown that with any LOP $2N$-port one can associate in a unique way
	two mathematical objects:
	\begin{itemize}
	\item a unitary matrix $\LOP{U}$ representing the LOP transformation:
	      $\bb{i} = \sum_{j}\LOP{U}_{ij}\ab{j}$;
	\item a unitary operator $U$ uniquely identified by the equations
	\begin{equation} \bb{i} = U^{\dag}\ab{i}U \quad i=1,\ldots, N\quad , \quad
	U|\mathbf{0}\rangle=|\mathbf{0}\rangle\, .
	\end{equation}
	\end{itemize}
	Moreover, we have shown that one can give an explicit procedure for building the operator $U$
	from the matrix $\LOP{U}$; conversely, if the operator $U$ is given, then the matrix
	$\LOP{U}$ can be obtained from relation (\ref{coincidenzaMatrici}).

	We conclude observing that the J-S map JS induces a map $\indotta \, :$ U($N$) $\rightarrow
	\mathcal{U}(\Hilb_{F}^{(N)})$
	where $\mathcal{U}(\Hilb_{F}^{(N)})$ is the group of unitary operators in $\Hilb_{F}^{(N)}$,
	defined by:
	\begin{equation}
	\indotta(\LOP{U})=\indotta (\mathrm{exp}(i\LOP{J})) = \mathrm{exp}(i \mathrm{JS}(\LOP{J})) = U\quad .
	\end{equation}
	Making use of the Stone-von Neumann theorem one can easily prove that
	\begin{equation}
	\indotta (\LOP{U}_{1}\LOP{U}_{2}) = \indotta (\LOP{U}_{1})\, \indotta (\LOP{U}_{2}) \quad ,
	\end{equation}
	 for any $\LOP{U}_{1},\LOP{U}_{2}\, \in\,$ U($N$), i.e. that $\indotta$ is a unitary representation of U($N$).

	\subsection{\label{sec2.4}
	            Basic examples}

	Now we give an explicit form to the objects we have introduced so far,
	 namely the matrix $\LOP{U}$ and the operator $U$,
	for two simple
	special cases: the 2- and the 4-port. They are indeed special because they
	are the only LOP directly implemented  in the labs respectively by
	\emph{phase shifters} (PS) and \emph{beam splitters} (BS);
	then, any generic linear optical multiport can be
	implemented decomposing it as an array of 2- and 4-ports ~\cite{rec}.

\begin{figure}[!h]
       \begin{center}
        \includegraphics[width=0.15\textwidth]{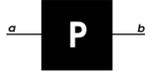}
        \caption{ \label{fig2} LOP $2$-port: phase-shifter.}
       \end{center}
      \end{figure}

	The simplest example is the PS, the LOP $2$-port (Fig.~\ref{fig2}) whose action
	is just the phase multiplication:
         \begin{equation}
          b = e^{+i\phi}a  \quad ,\quad
          b^{\dag} = e^{-i\phi}a^{\dag}\quad .
         \end{equation}

\begin{figure}[!h]
       \begin{center}
        \includegraphics[width=0.15\textwidth]{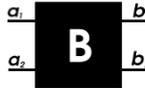}
        \caption{\label{fig3} LOP $4$-port: beam-splitter.}
       \end{center}
      \end{figure}

	The group involved is U(1), and its
	 generator is simply $\LOP{J}=1$; in this case, the J-S map associates with $\LOP{J}$
	the number operator, so that $J=a^{\dag}a$, and:
	     \begin{equation}
	      b = P^{\dag}\, a\, P  \quad \mathrm{with} \quad P(\phi)=e^{+i\phi \n}   \quad .
	     \end{equation}
	Notice that $P$ acts on the Fock space $\Hilb_{F}$ as a
	photon-number-dependent phase factor.
	\\
	A generic LOP $4$-port (Fig.~\ref{fig3}) is described
	by the $2\times 2$ unitary matrix $\LOP{B}$ ~\cite{yur,campos}:
	      \begin{equation}
	     \B= e^{i\phi_{0}}\B' = e^{i\phi_{0}}\left( \begin{array}{cc}
	            e^{i\phi_{\tau}}\cos\theta & e^{i\phi_{\rho}}\sin\theta \\
	            -e^{-i\phi_{\rho}}\sin\theta & e^{i\phi_{\tau}}\cos\theta
	                                \end{array} \right) \quad ;
	      \end{equation}
	 the  $4$-port is implemented by
        PSs and BSs, respectively corresponding to the U(1) factor
	and the SU(2)  matrix $B'$.

	To obtain the operator representation of the BS's SU(2) matrix,
	we consider the well-known Euler decomposition of a generic SU(2) matrix
	as the product of three elementary 'rotations'; then, recalling the
	bosonic realization  of the $\su{2}$ generators
	$J_{k}$, and the induced map $\indotta$, we can write:
	     \begin{equation}
	      B'(\alpha,\beta,\gamma) =   e^{-i\gamma J_{z}}e^{-i\beta J_{y}}e^{-i\alpha J_{z}} \quad ,
	     \end{equation}
	where $J_{i}=$JS$(\frac{1}{2}\sigma_{i})$ are the linear combinations of the operators $\{\ad{i}\ab{j}\}_{i,j=1}^{N}$
	given by equations (\ref{J}).
\section{\label{sec3}
         LOP schemes for deterministic quantum computation}
In this section we apply the results of
\S \ref{sec2.2}, \ref{sec2.3} and derive a constructive procedure for
building LOP circuits for deterministic quantum computation
on an arbitrary number of qubits; then we discuss the issue of scalability.

	\subsection{\label{sec3.1}
	         Encoding: the SPMQ scheme}
	The encoding of (quantum) information is generally made possible
	by a more or less strict correspondence between the mathematical properties
	of the symbols that are chosen to represent the information,
	and those of the description of the physical system that is chosen to encode symbols.
	The case of the SPMQ encoding is special from this point of view, since such a
	correspondence is indeed a complete equivalence in this case.
	\\
	As we have already said, symbols in quantum information are represented by qu$N$its; universal quantum computation
	can be done on strings of qubits in the common case of the binary coding.
	It is useful to recall the following abstract definitions:
	\begin{itemize}
	\item a \emph{quNit} is a vector in a $N$-dimensional abstract Hilbert space, endowed with the fundamental
		representation of U($N$) acting on it;
	\item a \emph{string of k  qubits}, or \emph{k-qubit}, is a vector in a $2^{k}$-dimensional Hilbert space ---
		specifically the tensor product of $k$ copies of the 2-dimensional single qubit Hilbert space ---
		endowed with the fundamental representation of U(2$^{k}$) acting on it.
	\end{itemize}
	It should be clear from the results of \S \ref{sec2.2} that the mathematical characterization of
	$\Hilb_{1}^{(N)}$ --- the space of the states of a single photon over $N$ optical modes ---
	 perfectly fits with the abstract definition of the qu$N$it space;
	furthermore, chosing $N=2^{k}$,
	one can say that $\Hilb_{1}^{(2^{k})}$ concides with the previous  abstract definition of the  Hilbert space of a string of $k$
	qubits.
	\\
	The SPMQ encoding is the one-to-one correspondence between \emph{logical states} --- i.e. the states of a string
	of k abstract qubits --- and the \emph{physical states} --- i.e. the states of the quantum system of a single photon
	over $2^{k}$ modes.
	The correspondence between logical and physical states can be formulated  explicitly,
	using te well known computational basis notation: if we denote by $\{|\mathtt{i}\rangle\}_{i=1}^{2^{k}}$
	the states of a given basis of the k logical qubits, we can rewrite them as
	column vectors with $2^{k}$ elements, that are all zero except for a 1 in the $i$-th position, with $i=1,\ldots ,2^{k}$.
	Then, the SPMQ scheme consists in encoding the logical  state $|\mathtt{i}\rangle$
	by the state of one photon in the $i$-th mode, with $i=1,\ldots ,2^{k}$,
	and the corresponding notation is:
	     \begin{eqnarray}\label{encode}
	      \mathrm{logical states} & \longleftrightarrow & \mathrm{physical states} \nonumber \\
					& & \nonumber \\
	      |\mathtt{i}\rangle \equiv \left(\begin{array}{c} 0\\ \vdots\\ 0\\ 1 \\ 0\\ \vdots\\ 0
		\end{array}\right)\} i & \longleftrightarrow &
	       |0\ldots 0\overbrace{1}^{i} 0 \ldots 0 \rangle = \ad{i}|\mathbf{0}\rangle \, .
	     \end{eqnarray}

	\subsection{\label{sec3.2}
			Elaboration: the algebraic scheme}
	We now claim that with this encoding scheme,
	LOP devices are
	sufficient to implement deterministically
	any quantum circuit, without the
	need for ancillary resources and
	postselection schemes ~\cite{klm1}.

	As we have previously shown, with any LOP $2N$-port one can associate a matrix in U($N$)
	acting on the input modes $\ab{1},\ldots ,\ab{n}$, or equivalently a unitary operator
	$U$ acting 'by similarity': $\ab{k}\,\mapsto\, U^{\dag}\ab{k}U$.
	From the physical point of view, this is nothing but the action of the time evolution operator
	associated with the $2N$-port (regarded as a quantum device) on the input field operators
	in the 'Heisenberg picture'.
	\\
	On the other hand, as far as applications to QIP and QC are concerned, since the encoding resource is given
	by the state vectors of the Fock space, it is convenient to switch to the 'Schr\''odinger picture'
	and to represent the action of the LOP devices (regarded as quantum logic gates)
	as the action of the associated unitary operators on the state vectors. The operator $U$ can be represented by an
	 infinite unitary matrix, after choosing a labeling of the Fock states; now we are interested
	in the action of $U$ on the subspace $\Hilb_{1}^{(N)}$, and, as shown in (\ref{primoBlocco}), this is
	represented by a U($N$) matrix $\LOPF{U}_{1}^{(N)}$ wich coincides with $\LOP{U}$.
	Since we know that any U($N$) matrix
	$\LOP{U}$ corresponds to a LOP
	$2N$-port built from an array of BS and PS ~\cite{rec},
	this means that we can act on
	$\Hilb_{1}^{(N)}$ with any desired
	unitary transformation, and so
	we can do any quantum computation on a
	string of a fixed number of qubits.
	\\
	To make the last statement more precise, once again we refer to
	the computational basis notation:
	recall that any logical quantum circuit acting on input strings of $k$ logical qubits
	is represented by some unitary operator $C$ acting on the $k$-qubit Hilbert space. After chosing a
	basis $\{|\mathtt{i}\rangle\}_{i=1}^{2^{k}}$ for the logical states, the operator $C$ associated with the circuit
	can be represented by a $2^{k}\times 2^{k}$ unitary matrix on such a basis:
	\begin{equation}
	(C)_{ij}\doteq\langle\mathtt{i}|C|\mathtt{j}\rangle\quad ;
	\end{equation}
	this is the \emph{computational basis matrix} of the quantum circuit.
	\\
	Then, when using the SPMQ encoding, in order to design the LOP circuit which implements a given $k$-qubit
	quantum circuit, one just needs to follow three simple steps:
	\begin{enumerate}
	\item  write down the computational
		basis matrix of the logical circuit;
	\item  take the $(C)_{ij}$ as the matrix elements of the
		$\LOPF{U}_{1}^{(N)}=\LOP{U}$ matrix of a LOP circuit;
	\item apply the RZBB procedure ˜\cite{rec} to decompose the $\LOP{U}$ matrix in the corresponding array of PS's and BS's.
	\end{enumerate}
	This simple, constructive procedure for designing LOP quantum circuits constitutes the demonstration
	of the claim we made at the beginning of this section.

	To conclude, notice that the simplicity of the procedure we presented
	makes it suitable for translation into an algorithm that could be run by a classical computer;
	furthermore,  if BS and PS with variable
	parameters were available,
	being their maximum number fixed by
	the number of modes $N$,
	LOP components
	could be rearranged automathically in
	the appropriate configuration, thus making the design of LOP quantum circuit a comletely authomatized process.

	\subsection{\label{sec3.3}
	         The issue of scalability}
	LOP circuits for deterministic quantum computation
	can be designed when encoding strings of qubits by
	single photon multimode states;
	but there are two practical problems related to this scheme.
	  \begin{enumerate}

		\item The first one is the fact that, in order to
		encode a $k$-qubit state, we need $2^{k}$ optical modes,
		which means that an exponential amount
		of physical resource is required;
		this limits the practical feasibility of
		circuits acting on an arbitrary number of qubits.

		\item On the other hand,
		this scheme is deterministic only for
		computations executed on a fixed number of qubits:
		when coupling two registers, physical states of
		2 photons will appear, which do not encode any logical state.
		Consider two strings of $k_{1}$ and $k_{2}$ qubits;
		these are encoded respectively on the spaces
		$\Hilb_{1}^{(N_{1})}$ and $\Hilb_{1}^{(N_{2})}$,
		where $N_{i}=2^{k_{i}}$. The resulting physical system
		after a generic LOP is the system of two photons on
		$N_{1}+N_{2}$ modes, whose Hilbert space
		$\Hilb_{2}^{(N_{1}+N_{2})}$
		is strictly larger than the encoding space
		$\Hilb_{1}^{(N_{1})}\otimes\Hilb_{1}^{(N_{2})}$.

     \end{enumerate}

	At present stage it is not
	clear yet what the architecture
	of a quantum computer will be, but it seems
	reasonable that it will be a hybrid object
	made out of different components; in this regard,
	scalability is only one of the requirements
	to be satisfied, and it should not be considered so
	stringent as to rule out a proposal for the implementation.
	With the scheme we presented here, one can build circuits
	acting on a small number, e.g. 2 or 3, of qubits,
	which allow to test experimentally with present technology
	 some interesting QIP protocols, as we show in the next section.
	As a further remark, we just point out a possible way
	to reduce the number $N$ of optical modes necessary to encode
	$k$-qubit logical states, in such a way that $N$ is polynomial in $k$.
	In fact, one could encode
	$k$-qubits in the subspace
	 $\Hilb_{n}^{(N)}$ of $\Hilb_{F}^{(N)}$, with
	$n\geq 2$, and suitably exploit the irreducible
	representations of U($N$) acting in such subspaces for
	implementing logic gates.

	With respect to the second problem, a solution
	could be found by means of a suitable
	postselected circuit that allows to reduce this
	deterministic non-scalable scheme to the non-deterministic
	scalable scheme proposed in ~\cite{klm1}, and \textit{viceversa}.

\section{\label{sec4}
         2-qubit deterministic circuits}

We first give some basic examples of 2-qubit gates,
and then a composite circuit
for the generation and measurement of
Bell states. To this aim,
we write down explicitly the linear map that encodes
logical 2-qubit states by single-photon
4-mode states, namely on $\Hilb_{1}^{(4)}$:
	\begin{equation}
		\begin{array}{ccccc}
		\mathrm{logical states} & {} & \mathrm{binary form} & {} & \mathrm{physical states}  \\
		|\mathtt{1}\rangle & \longleftrightarrow & |\mathtt{00}\rangle & \longleftrightarrow &  |1000\rangle \\
		|\mathtt{2}\rangle & \longleftrightarrow & |\mathtt{01}\rangle & \longleftrightarrow &  |0100\rangle \\ 
		|\mathtt{3}\rangle & \longleftrightarrow & |\mathtt{10}\rangle & \longleftrightarrow &  |0010\rangle \\ 
		|\mathtt{4}\rangle & \longleftrightarrow & |\mathtt{11}\rangle & \longleftrightarrow &  |0001\rangle \\
		\end{array}
	\end{equation}
where, following the conventional notation for $k$-qubit states, in the central column
 we have introduced the binary form of the logical state; we will always 
 use this notation in what follows.

	\subsection{\label{sec4.1}
	            cNOT}
\begin{figure}[!h]
	  \begin{center}
	   \includegraphics[width=0.3\textwidth]{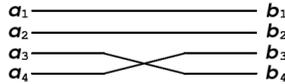}
	   \caption{\label{fig4} LOP cNOT.}
	  \end{center}
	 \end{figure}

	The first LOP circuit we present is a very simple one,
	implementing a cNOT gate. This is a 2-qubit
	universal gate,
	i.e. it can be shown that the cNOT
	and arbitrary 1-qubit gates are sufficient
	to build any quantum logic network.
	The cNOT gate acts on the logical computational
	basis states  flipping the second (\emph{target}) qubit
	when the first (\emph{control}) qubit is in
	the state $|\mathtt{1}\rangle$:
		\begin{eqnarray}
		 |\mathtt{00}\rangle  & \mapsto &  |\mathtt{00}\rangle \nonumber \\
		 |\mathtt{01}\rangle  & \mapsto &  |\mathtt{01}\rangle \nonumber \\
		 |\mathtt{10}\rangle  & \mapsto &  |\mathtt{11}\rangle \nonumber \\
		 |\mathtt{11}\rangle  & \mapsto &  |\mathtt{10}\rangle
		\end{eqnarray}
	and it is represented by the following matrix acting
	on the computational basis vectors:
		\begin{equation}
		 cNOT = \left(\begin{array}{cccc}
		              1&0&0&0\\
	                  0&1&0&0\\
		              0&0&0&1\\
		              0&0&1&0
		              \end{array}\right) .
		\end{equation}
	Exploiting the following matrix identity:
		\begin{equation}
		 cNOT = \LOPF{U}_{1}^{(4)} = \LOP{U}
		\end{equation}
	one can buid the LOP circuit  (Fig.~\ref{fig4})
	corresponding to $\LOP{U}$: decomposition ~\cite{rec}
	is trivial in this case,
	since $\LOP{U}$ itself describes the
	transformation of a BS with $100\% $
	transmission (that is, a simple exchange)
	coupling modes 3 and 4.
	Note that only one single photon source,
	three vacuum sources, and
	a classically controllable operation
	(the interchange of modes 3 and 4)
	are required, thus eliminating a possible
	source of errors due to non-ideal LOP components;
	the simplicity of this circuit is remarkable if one
	thinks that the cNOT is a basic gate that
	could be applied several times while
	running a quantum computation.

	\subsection{\label{sec4.2}
	            cPHASE}
	This gate append a desired phase facor $e^{-i\phi}$
	to the state $|\mathtt{11}\rangle$, while
	leaving the other unchanged;
	it can be shown by a simple calculation that
	a cNOT transformation can be obtained by a
	cPHASE with $\phi =\pi$ (also called cSIGN) preceeded
	and followed by a suitable transformation of the target qubit.

\begin{figure}[!h]
	  \begin{center}
	   \includegraphics[width=0.3\textwidth]{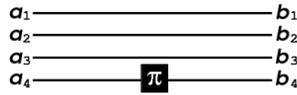}
	   \caption{\label{fig5} LOP cPHASE.}
	  \end{center}
	 \end{figure}

	The cPHASE is represented by the matrix:
		\begin{equation}
		 cPHASE = \left(\begin{array}{cccc}
		                1&0&0&0\\
		                0&1&0&0\\
		                0&0&1&0\\
		                0&0&0&e^{-i\phi}
		                \end{array}\right) \quad ;
		\end{equation}
	also in this case decomposition is trivial, and the
	corresponding LOP circuit requires only one PS with $\phi =\pi$
	acting on the 4-th mode,
	as depicted in Fig.~\ref{fig5}.

	\subsection{\label{sec4.3}
	            SWAP gate}
\begin{figure}[!h]
	  \begin{center}
	   \includegraphics[width=0.3\textwidth]{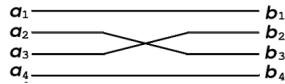}
	   \caption{\label{fig6} LOP SWAP.}
	  \end{center}
	 \end{figure}

	This is the gate that interchanges the logical state
	of the two qubits, represented by the matrix:
		\begin{equation}
		 SWAP = \left(\begin{array}{cccc}
					  1&0&0&0\\
					  0&0&1&0\\
					  0&1&0&0\\
		 			  0&0&0&1
		  			  \end{array}\right)
		\end{equation}
	and it correspond to a sequence of three alternate cNOT's.
	But within this scheme it is not necessary to
	implement this sequence: it suffices to note that
	the $SWAP$ matrix interpreted as a LOP matrix describes a
	BS with $100\%
	$ transmission coupling modes 2 and 3, which corresponds
	to the simple circuit shown in Fig.~\ref{fig6}.

	\subsection{\label{sec4.4}
	            Bell states production and analysis}
	Many QIP protocols, e.g. quantum teleportation
	~\cite{tele}, rely on the use of entangled states
	of qubits as a resource, and on the ability to
	distinguish among such states, thus leading to many efforts
	towards the production and detection of
	entangled  physical states.
	Bell states are defined as the four maximally
	entangled states of a 2-qubit system; within the scheme
	of QC they can be produced by means of a circuit
	composed by a cNOT preceeded by a Hadamard
	gate on the control qubit: this maps the logical
	computational basis states onto the Bell
	states:
		\begin{eqnarray}\label{bell}
		 |\mathtt{00}\rangle &  \mapsto &
		 \frac{|\mathtt{00}\rangle+|\mathtt{11}\rangle}{\sqrt{2}} \nonumber \\
		 |\mathtt{01}\rangle &  \mapsto &
		 \frac{|\mathtt{01}\rangle+|\mathtt{10}\rangle}{\sqrt{2}} \nonumber \\
		 |\mathtt{10}\rangle &  \mapsto &
		 \frac{|\mathtt{00}\rangle -|\mathtt{11}\rangle}{\sqrt{2}} \nonumber \\
		 |\mathtt{11}\rangle &  \mapsto &
	  	 \frac{|\mathtt{01}\rangle - |\mathtt{10}\rangle}{\sqrt{2}}
		\end{eqnarray}
	By a simple calculation one finds that
	it is represented by the followng matrix:
		\begin{equation}\label{bellmat}
		 cNOT\cdot(H_{1}\otimes\I_{2}) =
	     \left(\begin{array}{cccc}
		       \frac{1}{\sqrt{2}}&0&\frac{1}{\sqrt{2}}&0\\
				0&\frac{1}{\sqrt{2}}&0&\frac{1}{\sqrt{2}}\\
				0&\frac{1}{\sqrt{2}}&0&-\frac{1}{\sqrt{2}}\\
				\frac{1}{\sqrt{2}}&0&-\frac{1}{\sqrt{2}}&0
		       \end{array}\right)
		\end{equation}
	where $H$ denotes the Hadamard gate, and index 1
	refers to the fact that it acts on the first (control) qubit.

\begin{figure}[!h]
	  \begin{center}
	   \includegraphics[width=0.4\textwidth]{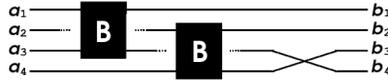}
	   \caption{\label{fig7} LOP circuit for Bell states
	   generation (running from left to right)
	   and measurement (from rigtht to left).
	   The two BS are balanced, and produce a
	   sign change on reflection off
	   the lower face.}
	  \end{center}
	 \end{figure}

	Implementation of a Bell state analyzer
	in the framework of linear optics
	has been studied ~\cite{calsa1,calsa2,calsa3}
	leading to the result
	that a complete measurement in the qubit
	polarization Bell basis is not
	possible. Nevertheless, in the SPMQ
	scheme we are proposing,
	a simple LOP circuit for the simulation of Bell
	state production and analysis can be found that
	works deterministically: as in the previous examples,
	one just takes the matrix (\ref{bellmat}) as 
	a LOP circuit matrix, decompose it
	and obtain the circuit depicted in Fig.~\ref{fig7}.
	Only two balanced ($50\%
	$ transmission) BS, with a sign change upon
	reflection off the lower side, and interchange of
	modes 3 and 4 are required.

\section{\label{sec5}
        Conclusions and perspectives}

	In this paper we have considered some algebraic structures of linear optics,
	and discussed how they provide a natural way to  deal with Linear-Optical
	Quantum Computing;  on the other hand, we have stressed on the correspondence
	of such algebraic objects with the basic instruments
	used in the laboratories, to point out that
	practical implementations can, in principle, be designed and tested.

	As a result, we first derived a SPMQ scheme
	for encoding any number of qubits
	by only one photon; then, we described an algorithmic procedure which,
	given any logical quantum circuit, allows to design the linear optical
	circuit which implements the logical circuit operation deterministically,
	and we presented some simple but interesting $2$-qubit gates
	and circuits.

	We also discussed some practical problems related to the scalability
	of such scheme, and pointed out some possible solutions.
	However it seems reasonable that in the future a
	quantum computer will be a composite object, which will exploit
	either scalable and non-scalable, either deterministic and
	non-deterministic components (most likely in association with classical components).

	In this regard, it is interesting to test such components
	experimentally, and further work is to be done to quantify the
	effects of non-ideal instruments on the theoretical schemes.
	It should be noticed that the scheme proposed here requires no
	ancillary systems, namely additional photon sources
	and counters, which are the main sources of inefficiency;
	\emph{only one} single photon source is required, regardless
	of the number of qubits, and this should lead to an efficiency
	which is almost independent on the number of qubits.



\section*{References}
\begin{thebibliography}{45}



\bibitem{Nielsen} See  Nielsen M and  Chuang I 2000
{\it ''Quantum Information and Quantum Computation''} (Cambridge University Press) and references therein.

\bibitem{klm1}  Knill E, Laflamme R and  Milburn G 2001 {\it Nature (London)}  \textbf{409} 46

\bibitem{scheel}  Scheel S,  Nemoto K, Munro W J, and Knight P L 2003 {\it Phys. Rev. } \textbf{A 68} 032310


\bibitem{lapaire}  Lapaire G G,  Kok P,  Dowling J P, and  Sipe J S 2003 {\it Phys. Rev. } \textbf{A 68} 042314


\bibitem{clausen}  Clausen J,  Knoll L, and  Welsch D G 2003 {\it Phys. Rev. } \textbf{A 68} 043822



   \bibitem{pjf1}  Pittman T B, Jacobs B C, and Franson J D 2001 {\it Phys. Rev.} \textbf{A 64} 062311

   \bibitem{ral1}  Ralph T C, White A G,  Munro W J,  and Milburn G J 2001 {\it Phys. Rev.} \textbf{A 65} 012314

   \bibitem{ral2}  Ralph T C, Langford N K, Bell T B,  and White A G 2002 {\it Phys. Rev.} \textbf{A 65} 062324

   \bibitem{ral3} Lund A P, and Ralph T C 2002  {\it Phys. Rev.} \textbf{A 66} 032307

   \bibitem{knill} Knill E 2002 {\it Phys. Rev.} \textbf{A 66} 052306

   \bibitem{ral4}  Dodd J L, Ralph T C, and Milburn G J 2003 {\it Phys. Rev.} \textbf{A 68} 042328

   \bibitem{giorgi} Giorgi G L, de Pasquale F, and Paganelli S 2004 {\it Phys. Rev.} \textbf{A 70} 022319


   \bibitem{pjf2} Pittman T B, Fitch M J, Jacobs B C, and  Franson J D 2003 {\it Phys. Rev.} \textbf{A 68} 032316

   \bibitem{san1}  Sanaka K \emph{et al.} 2003 {\it Nature} (London)  \textbf{421} 721-724

   \bibitem{obra}  O'Brien J L \emph{et al.} 2003 {\it Nature} (London)  \textbf{426} 264

   \bibitem{pjf3}  Pittman T B,  Jacobs B C, and Franson J D {\it Preprint} quant-ph/0404059

   \bibitem{gas} Gasparoni S \emph{et al.} {\it Preprint} quant-ph/0404107

   \bibitem{zhao}  Zhao Z \emph{et al.} {\it Preprint} quant-ph/0404129

   \bibitem{san2}  Sanaka K \emph{et al.} 2004 {\it Phys. Rev. Lett.} \textbf{92}(1)

   \bibitem{milburn} Milburn G J 1988 {\it Phys. Rev. Lett. }  \textbf{62} 2124

   \bibitem{cerf} Cerf N J, Adami C, and Kwiat P G 1998 {\it Phys. Rev.}  \textbf{A 57} R1477

   \bibitem{kurt} Englert B G, Kurtsiefer C, and  Weinfurter H 2003 {\it Phys. Rev.} \textbf{A 63} 032303

   \bibitem{fiore1}  Fiorentino M, Kim T, and  Wong F N C {\it Preprint}  quant-ph/0407136

   \bibitem{fiore2} Fiorentino M, and Wong F N C 2004 {\it Phys. Rev. Lett. }  \textbf{93} 070502


\bibitem{dixmier} This condition --- that appears in a remarkable paper by Dixmier J 1958
	{\it Comp. Math.} {\bf 13} 263-270 --- allows to rule out
	possible non-standard
	(i.e. inequivalent) realizations of  the canonical commutation
	relations; as an example of such a non-standard (but physically
	meaningful) realization, see
	Reeh H 1988 {\it J. Math. Phys.} {\bf  29} 1535-1536 .

\bibitem{stone}  Stone M H 1930 {\it Proc. Nat. Acad. Scie. U.S.A.} \textbf{16}
 172-175

\bibitem{von neumann}  von Neumann J 1931 {\it Math. Ann.} \textbf{104} 570-578


\bibitem{jordan} Jordan P 1935  {\it Z. Phys. } \textbf{94} 531

\bibitem{schw}  Schwinger J 1965 in {\it ''Quantum theory of angular momentum'' } L.C. Biedenharm
   and  H. Van Dam eds (Academic Press)

\bibitem{marmo}  Man'ko V I, Marmo G,  Vitale P, and Zaccaria F 1994 {\it Int. J. Mod. Phys. } \textbf{A9} 5541

\bibitem{wun} Wunsche A 2000 {\it Quantum Semiclass. Opt.} \textbf{2} 73-80

\bibitem{simon}  Simon B 1996 {\it "Representations of Finite and Compact Groups", Graduate Studies in Mathematics} \textbf{10} (AMS)



\bibitem{rec}  Reck M, Zeilinger A, Bernstein H J, and Bertani P 1994 {\it Phys. Rev. Lett.}   \textbf{73} 58

\bibitem{yur}  Yurke B, McCall S L, and Klauder J R 1986 {\it Phys. Rev.} \textbf{A 33} 4033

\bibitem{campos} Campos R A, Saleh B E A, and  Teich M C 1989 {\it Phys. Rev.} \textbf{A 40} 1371


\bibitem{calsa1} Lutkenhaus N,  Calsamiglia J, and  Suominen K A 1999 {\it Phys. Rev.} \textbf{A 59} 3295

\bibitem{calsa2}  Calsamiglia J and Lutkenhaus N 2001 {\it Appl. Phys. B} \textbf{59} 67

\bibitem{calsa3} Lutkenhaus N, Calsamiglia J, and Suominen K A 2002 {\it Phys. Rev.} \textbf{A 65} 030301(R)

\bibitem{tele} Bennett C H \emph{et al.} 1993 {\it Phys. Rev. Lett.} \textbf{70} 1895



\end{thebibliography}
\end{document}